\documentclass[submission,copyright,creativecommons]{eptcs}
\providecommand{\event}{PLACES 2017} 
\usepackage{underscore}           

\title{\parbox{\textwidth}{Towards a Categorical Representation of \newline Reversible Event Structures}}
\author{Eva Graversen \quad\quad\quad\quad\quad\quad Iain Phillips \quad\quad\quad\quad\quad\quad Nobuko Yoshida
\institute{Imperial College London, UK}
}
\def\titlerunning{Towards a Categorical Representation of Reversible Event Structures}
\def\authorrunning{E. Graversen, I. Phillips, \& N. Yoshida}
\usepackage[english]{babel}
\usepackage{amsmath}
\usepackage{amsfonts}
\usepackage{amssymb}
\usepackage{amsthm}
\usepackage{tikz}
\usepackage{stmaryrd}
\usepackage{mathtools}
\usepackage{multicol}
\usepackage{cite}
\usepackage{subcaption}
\usetikzlibrary{arrows.meta}
\newtheorem{theorem}{Theorem}[section]
\newtheorem{definition}[theorem]{Definition}

\newtheorem{proposition}[theorem]{Proposition}

\newtheorem{example}[theorem]{Example}
\title{Towards a Categorical Representation of\\ Reversible Event Structures}
\newcommand{\comment}[1]{}
\newcommand{\M}[1]{\mathcal{#1}}
\newcommand{\PES}[1]{(E_{#1},<_{#1},\cf_{#1})}
\newcommand{\AES}[1]{(E_{#1},<_{#1},\lhd_{#1})}
\newcommand{\RPES}[1]{(E_{#1},F_{#1},<_{#1},\cf_{#1},\prec_{#1},\rhd_{#1})}
\newcommand{\RAES}[1]{(E_{#1},F_{#1},\prec_{#1},\lhd_{#1})}
\newcommand{\RES}[1]{(E_{#1},\textsf{Con}_{#1},\vdash_{#1})}
\newcommand{\CS}[1]{(E_{#1},F_{#1},\textsf{C}_{#1},\rightarrow_{#1})}
\newcommand{\CStrans}[2]{\xrightarrow{#1\cup \underline{#2}}}
\newcommand{\csr}{R}
\newcommand{\Con}{\textsf{Con}}
\newcommand{\cat}[1]{\textbf{#1}}
\newcommand{\cf}{\mathrel{\sharp}}
\newcommand{\ob}{\mathrel{\obslash}}
\newcommand{\sdc}{\prec\!\!\prec}
\newcommand{\fin}{\mathrm{fin}}
\begin{document}
\maketitle
\begin{abstract}
We study categories for reversible computing, focussing on reversible
forms of event structures. Event structures are a well-established
model of true concurrency. There exist a number of forms of event
structures, including prime event structures, asymmetric event
structures, and general event structures. More recently, reversible
forms of these types of event structures have been defined. We formulate
corresponding categories and functors between them. We show that
products and co-products exist in many cases.
In most work on reversible computing, including reversible process
calculi, a cause-respecting condition is posited, meaning that the cause of an
event may not be reversed before the event itself. Since reversible
event structures are not assumed to be cause-respecting in general, we also define
cause-respecting subcategories of these event structures. Our longer-term aim is
to formulate event structure semantics for reversible process calculi.
\end{abstract}
\section{Introduction}\label{sec:intro}

Event structures \cite{nielsen1979petri}, a well-known model of true concurrency, consist of events and relations between them, describing the causes of events and conflict between events.
Winskel~\cite{winskel1982event} defined a category of event structures, and used this to define event structure semantics of CCS.

Reversible process calculi are a well-studied field \cite{lanese2014causal,danos2004reversible,DANOS200731,phillips2006reversing,cristescu2013compositional,lanese2010reversing}.
When considering the semantics of reversible processes, the ability to reverse events leads to finer distinctions of a true concurrency character~\cite{journals/entcs/PhillipsU07}; for example the CCS processes $a\mid b$ and $a.b+b.a$ can easily be distinguished by whether both $a$ and $b$ can be reversed at the end of the computation.
This motivates the study of \emph{reversible event structures}.
So far, no event structure semantics have been defined for reversible variants of CCS \cite{DANOS200731,danos2004reversible,phillips2006reversing}
(though the reversible $\pi$-calculus has been modelled using rigid families~\cite{CristescuKV16});
we intend this work to be one of the first steps towards doing so.

Reversible versions of various kinds of event structures were introduced
in~\cite{journals/jlp/PhillipsU15,phillips2013modelling}.
Our aim here is to interpret these as objects in appropriate categories and study functors between them.
So far few reversible frameworks have been defined categorically, though \cite{danos2007general} used category theory to describe the relationship between RCCS processes and their histories, and \cite{bowman2011dagger} used dagger categories to define a reversible process calculus called $\Pi$.

We define categories for the reversible event structures
from~\cite{journals/jlp/PhillipsU15,phillips2013modelling}, defining morphisms for each category and functors, and in some cases adjunctions, between them, along with coproducts, and, in the case of general reversible event structures, products.

With a few exceptions~\cite{phillips2012reversible,ulidowski2014concurrency},
reversible process calculi have always adopted \emph{causal} reversibility.
The reversible event structures of~\cite{journals/jlp/PhillipsU15,phillips2013modelling}
allow non-causal reversibility, inspired by bonding in biochemical processes.
We here define subcategories of the reversible event structures of~\cite{phillips2013modelling}
which are (1) \emph{stable},
meaning that the causes of an event cannot be ambiguous, which is clearly important for reversibility, and (2) \emph{cause-respecting}, meaning that no action can be reversed unless all the actions caused by it have been reversed first~\cite{journals/jlp/PhillipsU15}, which can be seen as a safety property for causal reversibility.
We show that under these conditions
any reachable configuration is forwards reachable (Theorem~\ref{the:fwdreach}).

We also consider configuration systems~\cite{journals/jlp/PhillipsU15}, a model of concurrency intended to serve a similar purpose as domains do for the forward-only event structures, letting the various kinds of reversible event structures be translated into one formalism. We show that, just as stable domains can be modelled as event structures, so finitely enabled configuration systems can be modelled as general reversible event structures, giving a tight correspondence in the stable setting (Theorem~\ref{the:CSRESinv}). 

\paragraph{Structure of the Paper.}
Section \ref{sec:forwards} reviews forwards-only event structures;
Section \ref{sec:RPES} looks at
reversible prime and asymmetric event structures,
while Section \ref{sec:RES} covers
reversible general event structures.
Section \ref{sec:CS} describes the category of configuration systems,
and Section \ref{sec:SRES} describes stable and cause-respecting
reversible event structures and configuration systems.



\section{Forwards-Only Event Structures}\label{sec:forwards}
\begin{figure}
	\center{\scalebox{0.85}{\begin{tikzpicture}
	\node (PES) at (-1,0) {\textbf{PES}};
	\node (Dom) at (3,-2) {\textbf{Dom}};
	\node (AES) at (3,0) {\textbf{AES}};
	\node (SES) at (6,0) {\textbf{SES}};
	\node (ES) at (9,0) {\textbf{ES}};
	
	\draw[-{Latex[length=1.5mm]}]
	(PES) edge[bend left=5] node[right] {$D_p$} (Dom)
	(PES) edge[bend left=10] node[auto] {$A$} node[below] {\rotatebox[origin=c]{90}{$\dashv$}} (AES)
	(Dom) edge[bend left=5] node[left] {$P_{pd}$} (PES)
	(AES) edge[bend left=10] node[auto] {$\Sigma$} (PES)
	(AES) edge node[auto] {$D_a$}(Dom)
	(PES) edge[bend left=20] node[auto] {$P_{ps}$} (SES)
	(SES) edge node[auto] {$I$} (ES)
	(SES) edge node[auto] {$D_s$} (Dom);
	\end{tikzpicture}}\caption{Categories of forward-only event structures and functors between them: PES were introduced in \cite{nielsen1979petri}, and defined categorically along with \cat{Dom}, \cat{SES}, \cat{ES}, $D_p$, $P_{pd}$, $P_{ps}$, and $D_s$ in \cite{winskel1986event}, \cat{AES}, $A$, and $D_a$ were introduced in \cite{baldan2001contextual}, and $\Sigma$ in \cite{journals/jlp/PhillipsU15}.
The adjunction between $A$ and $\Sigma$, denoted by $\dashv$, is new.}\label{fig:fwdcatfun}}
	
\end{figure}
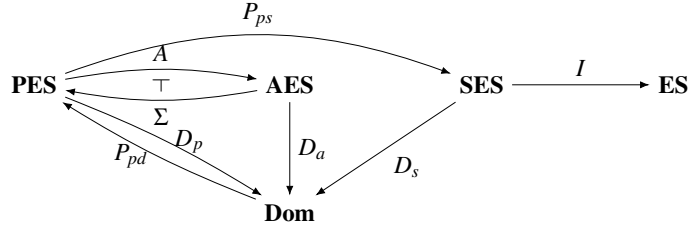

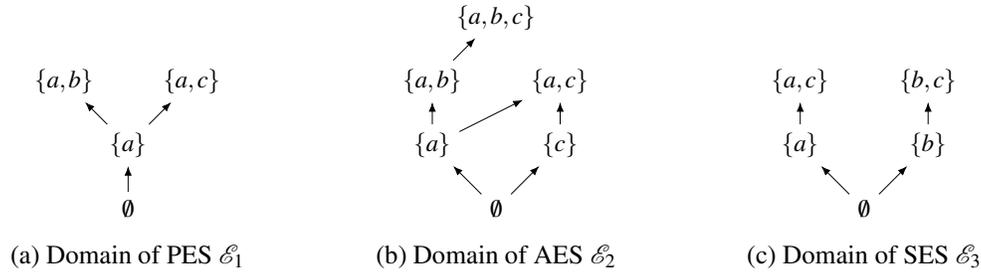
\begin{figure}
    \centering
    \begin{subfigure}[b]{0.3\textwidth}
        \center{\scalebox{0.85}{\begin{tikzpicture}
\node (e) at (0,0) {$\emptyset$};
\node (a) at (0,1) {$\{a\}$};
\node (ab) at (-1,2) {$\{a,b\}$};
\node (ac) at (1,2) {$\{a,c\}$};

\draw[-{Latex[length=1.5mm]}]
(e) edge (a)
(a) edge (ab)
(a) edge (ac);
\end{tikzpicture}}}
        \caption{Domain of PES $\M{E}_1$}
        \label{fig:PES}
    \end{subfigure}
    \begin{subfigure}[b]{0.3\textwidth}
        \center{\scalebox{0.85}{\begin{tikzpicture}
\node (em) at (0,0) {$\emptyset$};
\node (a) at (-1,1) {$\{a\}$};
\node (c) at (1,1) {$\{c\}$};
\node (ab) at (-1,2) {$\{a,b\}$};
\node (ac) at (1,2) {$\{a,c\}$};
\node (abc) at (0,3) {$\{a,b,c\}$};

\draw[-{Latex[length=1.5mm]}]
(em) edge (a)
(em) edge (c)
(a) edge (ab)
(a) edge (ac)
(c) edge (ac)
(ab) edge (abc)
;
\end{tikzpicture}}}
        \caption{Domain of AES $\M{E}_2$}
        \label{fig:AES}
    \end{subfigure}
    \begin{subfigure}[b]{0.3\textwidth}
        \center{\scalebox{0.85}{\begin{tikzpicture}
\node (em) at (0,0) {$\emptyset$};
\node (a) at (-1,1) {$\{a\}$};
\node (b) at (1,1) {$\{b\}$};
\node (ac) at (-1,2) {$\{a,c\}$};
\node (bc) at (1,2) {$\{b,c\}$};

\draw[-{Latex[length=1.5mm]}]
(em) edge (a)
(em) edge (b)
(a) edge (ac)
(b) edge (bc)
;
\end{tikzpicture}}}
        \caption{Domain of SES $\M{E}_3$}
        \label{fig:ES}
    \end{subfigure}
%
    \caption{Examples of domains representing event structures.}\label{fig:examples}
\end{figure}

Before describing the different categories of reversible event structures, we recall the categories of forward-only event structures and functors between them, as seen in Figure \ref{fig:fwdcatfun}.
A \emph{prime event structure} consists of a set of events, and causality and conflict relations describing when these events can occur. If $e<e'$ then $e'$ cannot happen unless $e$ has already happened. And if $e\cf e'$ then $e$ and $e'$ each prevent each other from occurring.
\begin{definition}[Prime Event Structure \cite{nielsen1979petri}]\label{def:PES}
A prime event structure (PES) is a triple $\M{E}=\PES{}$, where $E$ is the set of events and \emph{causality}, $<$, and \emph{conflict}, $\cf$, are binary relations on $E$ such that $\cf$ is irreflexive and symmetric, $<$ is an irreflexive partial order such that for every $e\in E$, $\{e'\mid e'< e\}$ is finite, and $\cf$ is hereditary with respect to $<$, i.e. for all $e,e',e''\in E$, if $e\cf e'$ and $e<e''$ then $e''\cf e'$.
\end{definition}
For any PES $\M{E}=\PES{}$, we say that $X\subseteq E$ is a configuration of $\M{E}$ if $X$ is left-closed under $<$ and conflict-free, meaning no $e,e'\in X$ exist, such that $e\cf e'$.
Configurations can be ordered by inclusion to form stable domains (coherent, prime algebraic, finitary partial orders)~\cite{winskel1986event}, as seen in Example \ref{ex:PEStoDom}.
\begin{example}\label{ex:PEStoDom}
The PES $\M{E}_1$ with events $a,b,c$ where $a<b$, $a<c$, and $c\cf b$, has configurations $\emptyset$, $\{a\}$, $\{a,b\}$, and $\{a,c\}$,
forming
the domain seen in Figure~\ref{fig:PES}.

\end{example}

Morphisms are defined on PESs in Definition \ref{def:PESmorph}, yielding the category \textbf{PES}. Morphisms on event structures act as a sort of synchronisation between the two structures, where if $X$ is a configuration then $f(X)$ is too, and two events, $e,e'$ can only synchronise with the same $f(e)=f(e')$ if they are in conflict.
\begin{definition}[PES morphism \cite{winskel1986event}]\label{def:PESmorph}
Let $\M{E}_0=(E_0,<_0,\cf_0)$ and $\M{E}_1=(E_1,<_1,\cf_1)$ be PESs. A morphism $f:\M{E}_0\rightarrow\M{E}_1$ is a partial function $f:E_0\rightarrow E_1$ such that  for all $e\in E_0$, if $f(e)\neq \bot$ then $\{e_1\mid e_1<_1f(e)\}\subseteq \{f(e')\mid e'<_0e\}$, and for all $e,e'\in E_0$, if $f(e)\neq\bot\neq f(e')$ and $f(e)\cf_1 f(e')$ or $f(e)=f(e')$ then $e\cf_0 e'$ or $e=e'$.
\end{definition}


%
%

\emph{Asymmetric event structures}~\cite{baldan2001contextual}\label{def:AES}
resemble prime event structures, with the difference being that the conflict relation $e\rhd e'$ (\cite{baldan2001contextual} uses the notation $e\nearrow e'$) is asymmetric, so that rather than $e$ and $e'$ being unable to coexist in a configuration, $e'$ cannot be added to a configuration that contains $e$. The converse relation $e'\lhd e$ can be seen as precedence or weak causation, where if both events are in a configuration then $e'$ was added first, as illustrated by Example \ref{ex:AEStoDom}.
%
%
An AES-morphism is defined in the same way as a PES morphism, but replacing symmetric conflict with asymmetric.
This gives the category \textbf{AES}. 
\begin{example}\label{ex:AEStoDom}
$\M{E}_2=\AES{}$ where $E=\{a,b,c\}$ and $a<b$ and $b\lhd c$ has configurations $\emptyset$, $\{a\}$, $\{c\}$, $\{a,b\}$, $\{a,c\}$, and $\{a,b,c\}$, and therefore $D_a(\M{E}_2)$ is the domain seen in Figure \ref{fig:AES}.
\end{example}

%


\emph{General} event structures, or simply \emph{event structures}, work somewhat differently from PESs or AESs. Instead of causation and conflict, they have an enabling relation and a consistency relation. 

\begin{definition}[Event structure \cite{winskel1986event}]\label{def:ES}
An event structure (ES) is a triple $\M{E}=\RES{}$, where $E$ is a set of events, $\Con\subseteq_{\fin} 2^E$ is the consistency relation, such that if $X\in \Con$ and $Y\subseteq X$ then $Y\in\Con$, and ${\vdash} \subseteq {\Con\times E}$ is the enabling relation, such that if $X\vdash e$ and $X\subseteq Y\in \Con$ then $Y\vdash e$.
\end{definition}
Configurations are finitely consistent sets of events,
where each event is deducible via the enabling relation.
\comment{
ES configurations are defined in Definition \ref{def:EScon}.

\begin{definition}[Configuration \cite{winskel1986event}]\label{def:EScon}
Given an ES $\M{E}=\RES{}$, a configuration of $\M{E}$ is a set $C\subseteq E$ such that
\begin{itemize}
\item for all $X\subseteq_{\fin} C$, $X\in \Con$
\item for all $e\in C$, there exists a sequence $e_0,e_1,\dots,e_n$ such that $e_n=e$ and for all $0\leq i\leq n$, $\{e_0,\dots,e_{i-1}\}\vdash e_i$
\end{itemize}

Again, the set of all configurations of $\M{E}$ is denoted $Conf(\M{E})$
\end{definition}
}
Once again we define an ES-morphism
, giving us the category \textbf{ES}~\cite{winskel1986event}.
The idea behind them is much the same as for PES- and AES-morphisms. Enabling sets are treated in much the same way as causes, and consistent sets in the opposite way from conflict.


\emph{Stable event structures}~\cite{winskel1986event}
form a full subcategory $\textbf{SES}$ of $\textbf{ES}$.
The idea is that in any given configuration, each event will have a unique enabling set.


\begin{example}\label{ex:ESex}$\M{E}_3=\RES{}$ where $E=\{a,b,c\}$, $\Con=\{\emptyset,\{a\},\{b\},\{a,c\},\{b,c\}\}$, and $\emptyset\vdash a$, $\emptyset \vdash b$, $\{a\}\vdash c$, and $\{b\}\vdash c$ can be represented by the domain $D_s(\M{E}_3)$ seen in Figure \ref{fig:ES}.
\end{example}


\comment{
AESs are not as easily mapped into ESs. Consider the AES $\M{E}=\AES{}$ where $E=\{a,b\}$ and $a\lhd b$. An attempt to describe $\M{E}$ as a RES $\RES{}$ where $E=\{a,b\}$ would require $\emptyset \vdash a$ and $\{b\}\in\Con$, but not $\{b\}\vdash a$, which is not possible. We have not found a way of mapping from \cat{AES} to \cat{ES}, while preserving the domain representation.
We shall see that this is possible in the reversible setting.
}



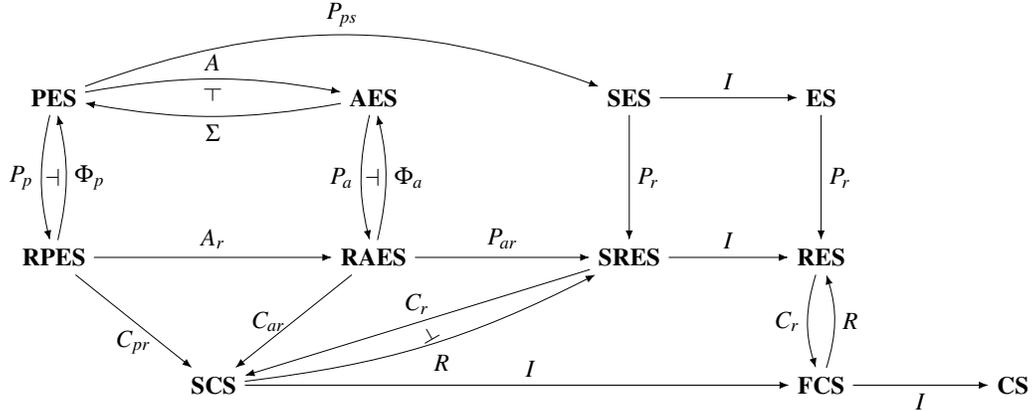
\begin{figure}
\center{\scalebox{0.85}{\begin{tikzpicture}
\node (PES) at (-3,0) {\textbf{PES}};
\node (AES) at (2,0) {\textbf{AES}};
\node (SES) at (6,0) {\textbf{SES}};
\node (ES) at (9,0) {\textbf{ES}};
\node (RPES) at (-3,-2.5) {\textbf{RPES}};
\node (RAES) at (2,-2.5) {\textbf{RAES}};
\node (SRES) at (6,-2.5) {\textbf{SRES}};
\node (RES) at (9,-2.5) {\textbf{RES}};
\node (SCS) at (-0.5,-4.5) {\textbf{SCS}};
\node (FCS) at (9,-4.5) {\textbf{FCS}};
\node (CS) at (12,-4.5) {\textbf{CS}};

\draw[-{Latex[length=1.5mm]}]
(PES) edge[bend right=13] node[left] {$P_p$} node[right=-0.08cm] {$\dashv$} (RPES)
(RPES) edge node[below] {$C_{pr}$}(SCS)
(PES) edge[bend left=10] node[auto] {$A$} node[below] {\rotatebox[origin=c]{90}{$\dashv$}} (AES)
(AES) edge[bend right=13] node[left] {$P_a$} node[right=-0.08cm] {$\dashv$} (RAES)
(RPES) edge node[auto] {$A_r$} (RAES)
(RAES) edge node[above] {$P_{ar}$} (SRES)
(RAES) edge node[left] {$C_{ar}$} (SCS)
(SRES) edge node[above] {$C_r$} node[below right=-0.12cm] {\rotatebox[origin=c]{300}{$\dashv$}} (SCS)
(RES) edge[bend right=17] node[left] {$C_r$} (FCS)
	(PES) edge[bend left=20] node[auto] {$P_{ps}$} (SES)
	(SES) edge node[auto] {$I$} (ES)
	(SRES) edge node[auto] {$I$} (RES)
	(SCS) edge node[above right] {$I$} (FCS)
	(ES) edge node[auto] {$P_r$} (RES)
	(SES) edge node[auto] {$P_r$} (SRES)
	;

\draw[{Latex[length=1.5mm]}-]
(PES) edge[bend left=13] node[right] {$\Phi_p$} (RPES)
(PES) edge[bend right=10] node[below] {$\Sigma$} (AES)
(AES) edge[bend left=13] node[right] {$\Phi_a$} (RAES)
(SRES) edge[bend left=10] node[auto] {$R$} (SCS)
(RES) edge[bend left=17] node[auto] {$R$} (FCS)
(CS) edge node[auto] {$I$} (FCS)
;
\end{tikzpicture}}\caption{Categories of event structures and functors between them: We extend Figure \ref{fig:fwdcatfun} by categorically defining RPESs, RAESs, CSs, $P_p$, $\Phi_p$, $P_a$, $\Phi_a$, $C_p$, $C_{pr}$, $C_a$, $C_{ar}$, and $A_r$ \cite{journals/jlp/PhillipsU15} and RESs and $P_r$ \cite{phillips2013modelling}.
The categories \cat{SRES}, \cat{SCS}, and \cat{FCS}, and functors $P_{pr}$, $P_{ar}$, $C_r$, $C$, and $R$ are new, as well as the noted adjunctions.}\label{fig:escatfun}}
\end{figure}
 
\section{Reversible Prime and Asymmetric Event Structures}\label{sec:RPES}
Our goal is to define the categories and functors in the lower part of Figure \ref{fig:escatfun}.

We start by adding reversibility to PESs. When discussing reversible events we will use $\underline{e}$ to denote reversing $e$ and $e^*$ to denote that $e$ may be performed or reversed. \emph{Reversible prime event structures}~\cite{journals/jlp/PhillipsU15}
(Definition~\ref{def:RPES}) consist of a set of events, $E$, some of which may be reversible, causality and conflict similar to a PES, reverse causality, which works similarly to causality, in that $e\prec \underline{e'}$ means $e'$ can only be reversed in configurations containing $e$, and prevention, which resembles the asymmetric conflict of AESs, in that $e\rhd \underline{e'}$ means that $e'$ can only be reversed in configurations not containing $e$.

\begin{definition}[RPES \cite{journals/jlp/PhillipsU15}]\label{def:RPES}
		A reversible prime event structure (RPES) is a sextuple $\M{E}=\RPES{}$ where $E$ is the set of events, $F\subseteq E$ is the set of reversible events, and
		\begin{itemize}
			\item $<$ is an irreflexive partial order such that for every $e\in E$, $\{ e'\in E\mid e'< e\}$ is finite and conflict-free 
			\item $\cf$ is irreflexive and symmetric such that if $e<e'$ then not $e\cf e'$
			\item $\rhd\subseteq E\times\underline{F}$ is the prevention relation
			\item ${\prec}\subseteq E\times \underline{F}$ is the reverse causality relation where for each $e\in F$, $e\prec \underline{e}$ and $\{e'\mid e'\prec\underline{e} \}$ is finite and conflict-free and if $e\prec \underline{e'}$ then not $e\rhd \underline{e'}$
			\item $\cf$ is hereditary with respect to sustained causation $\ll$ and $\ll$ is transitive, where $e\ll e'$ means that $e< e'$ and if $e\in F$ then $e'\rhd \underline{e}$
		\end{itemize}
	\end{definition}
	
As previously, in order to define the category \cat{RPES}, we need a notion of morphism. An RPES-morphism can be seen as a combination of a PES-morphism for the forwards part and an AES-morphism for the reverse part, and reversible events can only synchronise with other reversible events.
The category \cat{RPES} has coproducts
(Definition \ref{def:RPEScop}). A coproduct can be described as a choice between two event structures to behave as, as illustrated by Example \ref{ex:RPEScop}.

	\begin{definition}[RPES coproduct]\label{def:RPEScop}
Given RPESs $\M{E}_0=(E_0,F_0,<_0,\cf_0,\prec_0,\rhd_0)$ and $\M{E}_1=(E_1,F_0,<_1,\cf_1,\prec_1,\rhd_1)$, their coproduct $\M{E}_0+\M{E}_1$ is $(E,F,<,\cf,\prec,\rhd)$ where:
\begin{itemize}
\item $E=\{(0,e)\mid e\in E_0\}\cup \{(1,e)\mid e\in E_1\}$ and $F=\{(0,e)\mid e\in F_0\}\cup \{(1,e)\mid e\in F_1\}$
\item injection $i_j$ exist such that for $e\in E_j$, $i_j(e)=(j,e)$ for $j\in\{0,1\}$
\item $(j,e)<(j',e')$ iff $j=j'$ and $e<_je'$
\item $(j,e)\cf(j',e')$ iff $j\neq j'$ or $e\cf_je'$
\item $(j,e)\prec\underline{(j',e')}$ iff $j=j'$ and $e\prec_j \underline{e'}$
\item $(j,e)\rhd\underline{(j',e')}$ iff $e'\in F_{j'}$ and $j\neq j'$, or $e\rhd_j\underline{e'}$
\end{itemize}
\end{definition}

\begin{example}[RPES coproduct]\label{ex:RPEScop}
Given RPESs $\M{E}_0=\RPES{0}$ and $\M{E}_1=\RPES{1}$ where $E_0=\{a,b\}$, $F_0=\{a,b\}$, $a<_0 b$, $a\prec_0 \underline{b}$ and $E_1=\{c,d\}$, $F_1=\{c\}$, and $d\rhd_1 \underline{c}$, the coproduct $\M{E}_0+\M{E}_1$ is $\RPES{}$, where $E=\{ (0,a),(0,b),(1,c),(1,d)\}$, $F=\{ (0,a),(0,b),(1,c)\}$, $(0,a)<(0,b)$, $(0,a)\prec \underline{(0,b)}$, $(0,a)\cf(1,c)$, $(0,a)\cf (1,d)$, $(0,b)\cf (0,c)$, $(0,b)\cf (0,d)$, $(0,a)\rhd (1,c)$, $(0,b)\rhd (1,c)$, $(1,c)\rhd (0,a)$, $(1,d)\rhd (0,a)$, $(1,c)\rhd (0,b)$, $(1,d)\rhd (0,b)$, and $(1,d)\rhd \underline{(1,c)}$.
\end{example}

As we did with PESs, we will now add reversibility to AESs. \emph{Reversible asymmetric event structures} (RAES)~\cite{journals/jlp/PhillipsU15}
(Definition~\ref{def:RAES}) consist of events, some of which may be reversible, as well as causation and precedence, similar to an AES, except that $\prec$ is no longer a partial order, and instead just well-founded. In addition, both work on the reversible events, similarly to the RPES.
	
	\begin{definition}[RAES \cite{journals/jlp/PhillipsU15}]\label{def:RAES}
		A reversible asymmetric event structure (RAES) is a quadruple $\M{E}=\RAES{}$ where $E$ is the set of events, $F\subseteq E$ is the set of reversible events, and
		\begin{itemize}
			\item $\lhd\subseteq (E\cup \underline{F})\times E$ is the irreflexive precedence relation
			\item ${\prec} \subseteq E\times (E\cup\underline{F})$ is the causation relation, which is irreflexive and well-founded, such that for all $\alpha\in E\cup\underline{F}$, $\{e\in E\mid e\prec \alpha \}$ is finite and has no $\lhd$-cycles, and for all $e\in F$, $e\prec \underline{e}$
			\item for all $e\in E$ and $\alpha\in E\cup\underline{F}$ if $e\prec \alpha$ then not $e\rhd \alpha$
			\item $e \sdc e'$ implies $e\lhd e'$, where $e \sdc e'$ means that $e\prec e'$ and if $e\in F$ then $e'\rhd \underline{e}$
			\item $ \sdc$ is transitive and if $e\cf e'$ and $e \sdc e''$ then $e''\cf e'$
		\end{itemize}
	\end{definition}
	
Once again we create a category \cat{RAES} by defining RAES-morphisms. This definition is nearly identical to that of an AES-morphism, with the added condition that, like in the RPES morphism, reversible events can only synchronise with other reversible events.
The category \cat{RAES} has coproducts, defined
very similarly to the RPES coproduct, though without symmetric conflict and combining both causation relations into one.

\section{Reversible General Event Structures}\label{sec:RES}
The last kind of event structure we add reversibility to is the general event structure.
The \emph{reversible (general) event structure}
differs from
the general event structure, not only by allowing the reversal of events, but also by including a preventing set in the enabling relation, so that $X\ob Y\vdash e$ means $e$ is enabled in configurations that include all the events of $X$ but none of the events of $Y$. An example of an RES can be seen in Figure \ref{fig:CSasRES}. In all examples we will use $X\ob Y\vdash e^*$ as shorthand for $X' \ob Y \vdash e^*$ whenever $X\subseteq X'\in \Con$ 
\begin{definition}[RES \cite{phillips2013modelling}]\label{def:RES}
		A reversible event structure (RES) is a triple $\M{E}=\RES{}$ where $E$ is the set of events, $\Con\subseteq_{\fin} 2^E$ is the consistency relation, which is left-closed, ${\vdash} \subseteq {\Con\times 2^E\times (E\cup \underline{E})}$ is the enabling relation, and (1) if $X\ob Y\vdash e^*$ then $(X\cup \{e\})\cap Y=\emptyset$, (2) if $X\ob Y\vdash \underline{e}$ then $e\in X$, and (3) if $X\ob Y\vdash e^*$, $X\subseteq X'\in\textsf{Con}$, and $X'\cap Y=\emptyset$ then $X'\ob Y\vdash e^*$.
	\end{definition}
	

To define the category \cat{RES}, we need to define a RES-morphism (Definition \ref{def:RESmorph}). With the exception of the requirements regarding preventing sets, it is identical to the definition of an ES-morphism. We treat the preventing set similarly to (asymmetric) conflict in PES, AES, RPES, and RAES-morphisms.

\begin{definition}[RES morphism]\label{def:RESmorph}
Let $\M{E}_0=(E_0,\textsf{Con}_0,\vdash_0)$ and $\M{E}_1=(E_1,\textsf{Con}_1,\vdash_1)$ be RESs. A morphism $f:\M{E}_0\rightarrow \M{E}_1$ is a partial function $f:E_0\rightarrow E_1$ such that 
\begin{itemize}
\item for all $e\in E_0$, if $f(e)\neq \bot$ and $X\ob Y\vdash_0 e^*$ then there exists a $Y_1\subseteq E_1$ such that for all $e_0\in E_0$, if $f(e_0)\in Y_1$ then $e_0\in Y$ and $f(X)\ob Y_1\vdash_1 f(e)^*$
\item for any $X_0\in\textsf{Con}_0$, $f(X_0)\in \textsf{Con}_1$
\item for all $e,e'\in E_0 $, if $f(e)=f(e')\neq \bot$ and $e\neq e'$ then no $X\in \textsf{Con}_0$ exists such that $e,e'\in X$
\end{itemize} 
\end{definition}

As with \cat{RPES} and \cat{RAES}, \cat{RES} has coproducts (Definition \ref{def:REScop}).

\begin{definition}[RES coproduct]\label{def:REScop}
Given RESs $\M{E}_0=(E_0,\textsf{Con}_0,\vdash_0)$ and $\M{E}_1=(E_0,\textsf{Con}_0,\vdash_0)$, their coproduct $\M{E}_0+\M{E}_1$ is $(E,\textsf{Con},\vdash)$ where:
\begin{itemize}
\item $E=\{(0,e)\mid e\in E_0\}\cup \{(1,e)\mid e\in E_1\}$
\item injections $i_j$ exist such that for $e\in E_j$ $i_j(e)=(j,e)$ for $j\in\{0,1\}$
\item $X\in \textsf{Con}$ iff $\exists X_0\in \textsf{Con}_0. i_0(X_0)=X$ or $\exists X_1\in \textsf{Con}_1. i_1(X_1)=X$
\item $X\ob Y\vdash (j,e)^*$ iff $\exists X_j,Y_j\in E_j$ such that $X_j\ob Y_j\vdash e^*$, $i_j(X_j)=X$, $Y=i_j(Y_j)\cup (E\setminus i_j(E_j))$
\end{itemize}
\end{definition}

We also define the product of RESs (Definition \ref{def:RESpro}). A product can be described as a parallel composition of two RESs. The reason we did not define the products of RPESs or RAESs is, that while the ES product defined in \cite{winskel1986event} easily translates to RESs, definitions of PES products, such as the one based on mapping the PESs into a domain and back seen in \cite{vaandrager1989simple}, are far more complex and difficult to translate directly to a reversible setting. Since we do not have mappings from CSs to RPESs or RAESs, this is not a possible solution. Example \ref{ex:RESpro} shows the product of two RESs.

\begin{definition}[RES product]\label{def:RESpro}
Given RESs $\M{E}_0=(E_0,\textsf{Con}_0,\vdash_0)$ and $\M{E}_1=(E_1,\textsf{Con}_1,\vdash_1)$, their partially synchronous product $\M{E}_0\times \M{E}_1$ is $(E,\textsf{Con},\vdash)$ where:
\begin{itemize}
\item $E=E_0\times_* E_1=\{(e,*) \mid e\in E_0\} \cup \{(*,e) \mid e\in E_1\} \cup \{(e,e') \mid e\in E_0\text{ and } e'\in E_1\}$
\item there exist projections $\pi_0,\pi_1$ such that for $(e_0,e_1)\in E$, $\pi_i((e_0,e_1))=e_i$
\item $X\in \textsf{Con}$ if $\pi_0(X)\in \textsf{Con}_0$, $\pi_1(X)\in \textsf{Con}_1$, and for all $e,e'\in X$, if $\pi_0(e)=\pi_0(e')$ or  $\pi_1(e)=\pi_1(e')$ then $e=e'$
\item $X\ob Y\vdash e^*$ if
\begin{itemize}
\item if $\pi_0(e)\neq*$ then $\pi_0(X)\ob \pi_0(Y)\vdash \pi_0(e)^*$
\item if $\pi_1(e)\neq *$ then $\pi_1(X)\ob \pi_1(Y)\vdash \pi_1(e)^*$
\item if $e^*=\underline{e}$ then $e\in X$
\end{itemize}
\end{itemize}
\end{definition}

\begin{example}[RES product]\label{ex:RESpro}
Given RESs $\M{E}_0=\RES{0}$ and $\M{E}_1=\RES{1}$, where $E_0=\{a,b\}$, $\Con_0=2^{E_0}$, $\emptyset\ob \emptyset \vdash_0 a$, $\{a\}\ob \emptyset \vdash_0 b$, $\{a,b\}\ob \emptyset \vdash_0 \underline{b}$, and $\{a\}\ob \emptyset \vdash_0 \underline{a}$ and $E_1=\{c\}$, $\Con_1=\{\emptyset,\{c\}\}$, $\emptyset \ob \emptyset \vdash_1 c$, and $\{c\}\ob \vdash_1 \underline{c}$, the product $\M{E}_0\times \M{E}_1$ is $\RES{}$ where $E=\{(a,*),(b,*),(a,c),\;\;\;\;$ $(b,c),(*,c)\}$, $\;\Con=\{\emptyset, \{(a,*)\},\{(b,*)\},\{(a,c)\}, \{(b,c)\},\, \{(*,c)\}, \{(a,*),(b,*)\}, \{(a,*),(b,c)\}, \; \; \;\;\; \;$ $\{(a,*),(*,c)\},\{(a,c),(b,*)\}, \{(b,*),(*,c)\}, $ $\{(a,*),(b,*),(*,c)\}\}$, $\emptyset \ob \emptyset \vdash (a,*)$, $\{(a,*)\ob \emptyset \vdash (b,*)$, $\{(a,c)\}\ob \emptyset \vdash (b,*)$, $\emptyset \ob \emptyset \vdash (a,c)$, $\{(a,*)\ob \emptyset \vdash (b,c)$, $\{(a,c)\}\ob \emptyset \vdash (b,c)$, $\emptyset\ob\emptyset\vdash (*,c)$, $\{(a,*)\}\ob \emptyset \vdash \underline{(a,*)}$, $\{(b,*),(a,*)\}\ob \emptyset \vdash \underline{(b,*)}$, $\{(b,*),(a,c)\}\ob \emptyset \vdash \underline{(b,*)}$, $\{(a,c)\}\ob \emptyset \vdash \underline{(a,c)}$, $\{(b,c),(a,*)\}\ob \emptyset \vdash \underline{(b,c)}$, and $\{(*,c)\}\ob \emptyset \vdash \underline{(*,c)}$.
\end{example}

We also create functors from \cat{RPES} and \cat{RAES} to \cat{RES}. While not all AESs have ESs which map to the same domain, RAESs map into RESs using the preventing set to model asymmetric conflict as described in Definition \ref{def:RAEStoRESfun}.  
\begin{definition}[From \cat{RAES} to \cat{RES}]\label{def:RAEStoRESfun} 
The mapping $P_{ar}:\cat{RAES}\rightarrow \cat{RES}$ is defined as:
\begin{itemize}
\item $P_{ar}(\M{E})=(E,\textsf{Con},\vdash)$ where

\textsf{Con}$=\{X\subseteq E\mid \lhd \text{ is well-founded on }X\}$

$X\ob Y\vdash e$ if $\{ e' \mid e'\prec e\}\subseteq X\in\textsf{Con} $, $Y= \{e'\mid e'\rhd e\}$, $X\cap Y=\emptyset$, and $e\in E$

$X\ob Y\vdash \underline{e}$ if $ \{ e' \mid e'\prec \underline{e}\}\subseteq X\in \textsf{Con}$, $Y= \{e'\mid e'\rhd \underline{e}\}$, $X\cap Y=\emptyset$, and $e\in F$
\item $P_{ar}(f)=f$
\end{itemize}

\end{definition}

\section{Configuration Systems}\label{sec:CS}
\begin{figure}
\begin{subfigure}[b]{0.45\textwidth}
    \center{\scalebox{0.85}{\begin{tikzpicture}
\node (em) at (0,0) {$\emptyset$};
\node (a) at (-1,1) {$\{a\}$};
\node (b) at (1,1) {$\{b\}$};
\node (ab) at (0,2) {$\{a,b\}$};

\draw[-{Latex[length=1.5mm]}]
(em) edge (a)
(em) edge (b)
(a) edge (ab)
(b) edge (ab)
(em) edge (ab)
(b) edge (em)
(b) edge (a)
(ab) edge (a)
(a) edge (em)
;
\end{tikzpicture}}}
\caption{CS $\M{C}$}
\label{fig:CS}
    \end{subfigure}
\begin{subfigure}[b]{0.45\textwidth}
\scalebox{0.85}{\begin{tabular}{l}
$\M{E}=\RES{}$ where \\
$E=\{a,b\}$ \\
$\Con=\{\emptyset,\{a\},\{b\},\{a,b\}\}$ \\
$\emptyset\ob\emptyset \vdash a$, $\emptyset\ob\emptyset \vdash b$, \\
$\{b\}\ob\emptyset \vdash \underline{b}$, $\{a\}\ob\{b\} \vdash \underline{a}$
\end{tabular}}
\caption{RES $\M{E}$}
\label{fig:CSasRES}
\end{subfigure}
\caption{A CS and the corresponding RES such that $R(\M{C})=\M{E}$ and $C_r(\M{E})=\M{C}$.}
\label{fig:CStoRES}
\end{figure}
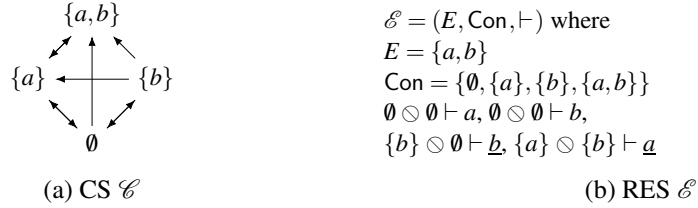

Configuration systems perform a similar role in the reversible setting to domains in the forward-only setting,
though they have a more operational character. A configuration system~\cite{journals/jlp/PhillipsU15} (Definition~\ref{def:CS}) consists of a set of events, $E$, some of which, $F$, are reversible, a set $\textsf{C}$ of configurations on these, and an optionally labelled transition relation $\rightarrow$ such that if $X\CStrans{A}{B} Y$ then the events of $A$ can happen and the events of $B$ can be undone in any order starting from configuration $X$, resulting in $Y$. We also leave out $Y$ when describing such a transition, since it is implied that $Y=(X\setminus B)\cup A$. A CS is shown in Figure \ref{fig:CS}.

\begin{definition}[Configuration system \cite{journals/jlp/PhillipsU15}]\label{def:CS}
A configuration system (CS) is a quadruple $\M{C} = \CS{}$ where $E$ is a set of events, $F\subseteq E$ is a set of reversible events, $\textsf{C}\subseteq 2^E$ is the set of configurations, and  ${\rightarrow} \subseteq {\textsf{C}\times 2^{E\cup\underline{F}} \times \textsf{C}}$ is an optionally labelled transition relation such that if $X\CStrans{A}{B} Y$ then:
\begin{itemize}
\item $A\cap X=\emptyset$, $B\subseteq X\cap F$, and $Y=(X\setminus B)\cup A$
\item for all $A'\subseteq A$ and $B'\subseteq B$, we have $X\CStrans{A'}{B'} Z \CStrans{(A\setminus A')}{(B\setminus B')}Y$, meaning $Z=(X\setminus B')\cup A'\in \textsf{C}$
\end{itemize}
\end{definition} 

%


We define a notion of morphism (Definition \ref{def:CSmorph}), creating the category \cat{CS}.

\begin{definition}[CS-morphism]\label{def:CSmorph}
Let $\M{C}_0=\CS{0}$ and $\M{C}_1=\CS{1}$ be configuration systems. A configuration system morphism is a partial function $f:E_0\rightarrow E_1$ such that
\begin{itemize}
\item for any $X,Y\in \textsf{C}_0$, $A\subseteq E_0$, and $B\subseteq F_0$, if $X \xrightarrow{A \cup \underline{B}}_0 Y$ then $f(X)\xrightarrow{f(A)\cup f(\underline{B})}_1 f(Y)$
\item for any $X\in \textsf{C}_0$, $f(X)\in \textsf{C}_1$
\item for all $e_0,e_0'\in E_0$, if $f(e_0)=f(e_0')\neq \bot$ and $e_0\neq e_0'$ then there exists no $X\in \textsf{C}_0$ such that $e_0,e_0'\in X$ 
\end{itemize}
\end{definition}

We also define the coproduct of two CSs (Definition~\ref{def:CScop}). This is illustrated with CSs modelling the RPESs and RESs from Examples \ref{ex:RPEScop} and \ref{ex:RESpro} in Example~\ref{ex:CScop}.

\begin{definition}[CS coproduct]\label{def:CScop}
Given CSs $\M{C}_0=\CS{0}$ and $\M{C}_1=\CS{1}$, their coproduct $\M{C}_0+\M{C}_1=\CS{}$ where:
\begin{itemize}
\item $E=\{(0,e)\mid e\in E_0\}\cup \{(1,e)\mid e\in E_1\}$ and $F=\{(0,e)\mid e\in F_0\}\cup \{(1,e)\mid e\in F_1\}$
\item injections $i_j$ exist such that for $e\in E_j$ $i_j(e)=(j,e)$ for $j\in\{0,1\}$
\item $X\in \textsf{C}$ iff $\exists X_0\in \textsf{C}_0. i_0(X_0)=X$ or $\exists X_1\in \textsf{C}_1. i_1(X_1)=X$
\item $X\CStrans{A}{B} Y$ iff there exists $j\in \{0,1\}$ such that there exist $X_j,Y_j,A_j,B_j\subseteq E_j$ such that $i_j(X_j)=X$, $i_j(Y_j)=Y$, $i_j(A_j)=A$, $i_j(B_j)=B$, and $X_j \CStrans{A_j}{B_j}_j Y_j$.
\end{itemize}
\end{definition}


\begin{example}[Coproduct]\label{ex:CScop}\mbox{}
\vspace*{-17pt}
\center{\hspace*{1cm}\scalebox{0.85}{\begin{tikzpicture}
\node (em1) at (1,0) {$\emptyset$};
\node (a) at (1,1) {$\{a\}$};
\node (b) at (0,1) {$\{b\}$};
\node (ab) at (1,2) {$\{a,b\}$};
\node (plus) at (2.5,0.5) {{\huge $+$}};
\node (em2) at (4.5,0) {$\emptyset$};
\node (c) at (4,1) {$\{c\}$};
\node (d) at (5,1) {$\{d\}$};
\node (cd) at (4.5,2) {$\{c,d\}$};
\node (eq) at (6,0.5) {{\huge $=$}};
\node (em3) at (8.5,0) {$\emptyset$};
\node (a3) at (8,1) {$\{a\}$};
\node (b3) at (7,1) {$\{b\}$};
\node (ab3) at (8,2) {$\{a,b\}$};
\node (c3) at (9,1) {$\{c\}$};
\node (d3) at (10,1) {$\{d\}$};
\node (cd3) at (9.5,2) {$\{c,d\}$};

\draw[-{Latex[length=1.5mm]}]
(em1) edge (a)
(a) edge (ab)
(ab) edge (a)
(b) edge (ab)
(ab) edge (b)
(em2) edge (c)
(c) edge (em2)
(em2) edge (d)
(c) edge (cd)
(d) edge (cd)
(em2) edge (cd)
(em3) edge (a3)
(a3) edge (ab3)
(ab3) edge (a3)
(b3) edge (ab3)
(ab3) edge (b3)
(em3) edge (c3)
(c3) edge (em3)
(em3) edge (d3)
(c3) edge (cd3)
(d3) edge (cd3)
(em3) edge[bend right=25] (cd3)
;
\end{tikzpicture}}}
\end{example}

We also define the product of CSs (Definition~\ref{def:CSpro}). This is illustrated in Example \ref{ex:CSpro}, where the CSs represent the RESs of Example \ref{ex:RESpro}.

\begin{definition}[CS product]\label{def:CSpro}
Given CSs $\M{C}_0=\CS{0}$ and $\M{C}_1=\CS{1}$, their partially synchronous product $\M{C}_0\times \M{C}_1=\CS{}$ where:
\begin{itemize}
\item $E=E_0\times_* E_1=\{(e,*) \mid e\in E_0\} \cup \{(*,e) \mid e\in E_1\} \cup \{(e,e') \mid e\in E_0\text{ and } e'\in E_1\}$
\item $F=F_0\times_* F_1=\{(e,*) \mid e\in F_0\} \cup \{(*,e) \mid e\in F_1\} \cup \{(e,e') \mid e\in F_0\text{ and } e'\in F_1\}$
\item there exist projections $\pi_0,\pi_1$ such that for $(e_0,e_1)\in E$, $\pi_i((e_0,e_1))=e_i$
\item $X\in \textsf{C}$ if $\pi_0(X)\in \textsf{C}_0$, $\pi_1(X)\in \textsf{C}_1$, and for all $e,e'\in X$, if $\pi_0(e)=\pi_0(e')$ or $\pi_1(e)=\pi_1(e')$ then $e=e'$
\item $X\CStrans{A}{B} Y$ if $B\subseteq X$ and
\begin{itemize}
\item if $\pi_0(A\cup B)\neq\emptyset$ then $\pi_0(X)\CStrans{\pi_0(A)}{\pi_0(B)}_0 \pi_0(Y)$
\item if $\pi_1(A\cup B)\neq\emptyset$ then $\pi_1(X)\CStrans{\pi_1(A)}{\pi_1(B)}_1 \pi_1(Y)$
\end{itemize}
\end{itemize}
\end{definition}
\begin{example}[Product]\label{ex:CSpro}\mbox{}
\vspace*{-17pt}
\center{\scalebox{0.85}{\begin{tikzpicture}
\node (em1) at (0.25,0) {$\emptyset$};
\node (a) at (0.25,1) {$\{a\}$};
\node (b) at (-.75,1) {$\{b\}$};
\node (ab) at (0.25,2) {$\{a,b\}$};
\node (plus) at (1,0.5) {{\Large $\times$}};
\node (em2) at (1.75,0) {$\emptyset$};
\node (c) at (1.75,1) {$\{c\}$};
\node (eq) at (2.5,0.5) {{\Large $=$}};
\node (em3) at (9.5,0) {$\emptyset$};
\node (a*) at (11.5,1) {$\{(a,*)\}$};
\node (b*) at (7.5,1) {$\{(b,*)\}$};
\node (ac) at (5.5,1) {$\{(a,c)\}$};
\node (bc) at (13.5,1) {$\{(b,c)\}$};
\node (*c) at (9.5,1) {$\{(*,c)\}$};
\node (a*b*) at (9.5,2) {$\{(a,*),(b,*)\}$};
\node (a*bc) at (14.5,2) {$\{(a,*),(b,c)\}$};
\node (a**c) at (12,2) {$\{(a,*),(*,c)\}$};
\node (acb*) at (4.5,2) {$\{(a,c),(b,*)\}$};
\node (b**c) at (7,2) {$\{(b,*),(*,c)\}$};
\node (a*b**c) at (9.5,3) {$\{(a,*),(b,*),(*,c)\}$};

\draw[-{Latex[length=1.5mm]}]
(em1) edge (a)
(a) edge (ab)
(ab) edge (a)
(b) edge (ab)
(ab) edge (b)
(em2) edge (c)
(c) edge (em2)
(em3) edge (a*)
(em3) edge (ac)
(em3) edge (*c)
(*c) edge (em3)
(a*) edge (a*b*)
(a*b*) edge (a*)
(a*) edge (a*bc)
(a*bc) edge (a*)
(a*) edge (a**c)
(a**c) edge (a*)
(b*) edge (a*b*)
(a*b*) edge (b*)
(b*) edge (acb*)
(acb*) edge (b*)
(b*) edge (b**c)
(b**c) edge (b*)
(ac) edge (acb*)
(acb*) edge (ac)
(bc) edge (a*bc)
(a*bc) edge (bc)
(*c) edge (a**c)
(a*b*) edge (a*b**c)
(a*b**c) edge (a*b*)
(a**c) edge (a*b**c)
(a*b**c) edge (a**c)
(b**c) edge (a*b**c)
(a*b**c) edge (b**c)
(em3) edge (a**c)
(b*) edge[bend left=12] (a*b**c)
(a*b**c) edge[bend right=12] (b*)
(a*) edge[bend right=12] (a*b**c)
(a*b**c) edge[bend left=12] (a*)
;
\end{tikzpicture}}}

\end{example}

We define a functor $C_r$ from \cat{RES} to \cat{CS} (Definition \ref{def:REStoCSfun}).

\begin{definition}[From \textbf{RES} to \textbf{CS}]\label{def:REStoCSfun}
The mapping $C_{r}:\textbf{RES} \rightarrow \textbf{CS}$ is defined as 
\begin{itemize}
\item $C_{r}(\RES{})=(E,F,\textsf{C},\rightarrow)$, where (1) $e\in F$ if there exists $X,Y$ such that $X\ob Y\vdash \underline{e}$, (2) $C\in \textsf{C}$ if for all $X\subseteq_{\fin} C$, $X\in \Con$, and (3) for $X,Y\in \textsf{C}$, $X\CStrans{A}{B}Y$ if
\begin{itemize}
\item $Y=(X\setminus B)\cup A$, $A\cap X=\emptyset$, $B\subseteq X$, and $X\cup A\in \textsf{C}$ 
\item for all $e$ in $A$, $X'\ob Z \vdash e$ for some $X',Z$ such that $X'\subseteq_{\fin} X\setminus B$ and $Z\cap(X\cup A)=\emptyset$
\item for all $e\in B$, $X'\ob Z \vdash \underline{e}$ for some $X',Z$ such that $X'\subseteq_{\fin} X\setminus (B\setminus\{e\})$ and $Z\cap (X\cup A)=\emptyset$
\end{itemize}
\item $C_{r}(f)=f$
\end{itemize}
\end{definition}

Applying this functor to a RES results in a \emph{finitely enabled CS} (FCS), that is to say a CS such that there does not exist a transition from an infinite configuration $X\CStrans{A}{B}$, such that there does not exist a finite configuration $X'\subseteq_{\fin} X$ such that $X'\CStrans{A}{B}$ and whenever $X'\subseteq X''\subseteq X$ there exists a transition $X''\CStrans{A}{B}$. We call the category of these CSs and the CS-morphisms between them \cat{FCS}, and describe a functor, $\csr$, from this category to \cat{RES} in Definition \ref{def:CStoRESfun}. An example of $C_r$ and $R$ can be seen in Figure \ref{fig:CStoRES}.
\begin{definition}[From \cat{FCS} to \cat{RES}]\label{def:CStoRESfun} 
The mapping $\csr:\textbf{FCS}\rightarrow \textbf{RES}$ is defined as:
\begin{itemize}
\item $\csr(\CS{})=\RES{}$ where $X\in\textsf{Con}$ if $X\subseteq_\fin C\in\textsf{C}$ and: 
\begin{itemize}
\item If $X\xrightarrow{\{e^*\}}$ and 
\begin{itemize}
\item $X'\subseteq X$, $X'\xrightarrow{\{e^*\}}$, and whenever $X'\subseteq X''\subseteq X$ there exists a transition $X''\xrightarrow{\{e^*\}}$
\item no $X''\subset X'$ exists such that $X''\xrightarrow{\{e^*\}}$, and whenever $X''\subseteq X'''\subseteq X$ there exists a transition $X'''\xrightarrow{\{e^*\}}$
\item no $X''\supset X$ exists such that $X''\xrightarrow{\{e^*\}}$, and whenever $X'\subseteq X'''\subseteq X''$ there exists a transition $X'''\xrightarrow{\{e^*\}}$
\end{itemize}
then 
\begin{itemize}
\item if $e^*=e$, then for all $X''\in \Con$ such that $X'\subseteq X''\subseteq X\cup \{e\}$, $X''\ob E\setminus X\cup \{e\} \vdash e$
\item if $e^*=\underline{e}$, then for all $X''\in \Con$ such that $X'\subseteq X''\subseteq X$, $X''\ob E\setminus (X\setminus \{e\} \vdash \underline{e}$
\end{itemize}
\end{itemize}
\item $\csr(f)=f$
\end{itemize}
\end{definition}

As Theorem~\ref{the:CSRESinv} states, $C_r$ and $R$ are in many cases inverses of each other. 

\section{Stable Reversible Event Structures and Configuration Systems}\label{sec:SRES}

Similarly to the stable event structures, we define the \emph{stable reversible event structures} (Definition~\ref{def:SRES}), and create the category \cat{SRES} consisting of SRESs and the RES-morphisms between them. SRESs and SESs are defined identically, with the exception that in an SRES the preventing sets are included as well, and treated in much the same way as the enabling sets. Like in a SES, an event in a configuration of a SRES will have one possible cause as long as the configuration has been reached by only going forwards. 
\begin{definition}[Stable RES]\label{def:SRES}
A stable reversible event structure (SRES) is an RES $\M{E}=(E,\textsf{Con},\vdash)$ such that for all $e^*\in E$ if $X\ob Y\vdash e^*$, $X'\ob Y'\vdash e^*$, and $X\cup X'+e^*\in \textsf{Con}$ then ${X\cap X'} \ob {Y\cap Y'} \vdash e^*$.
\end{definition}

Similarly, we can define a \emph{stable configuration system} (Definition \ref{def:SCS}). This has the property that if $\M{E}$ is a SRES then $C_r(\M{E})$ is a SCS, and if $\M{C}$ is a SCS then $R(\M{C})$ is a SRES.

\begin{definition}[Stable CS]\label{def:SCS}
A stable CS (SCS) is an FCS $\M{C}=\CS{}$ such that 
\begin{enumerate}
\item $\textsf{C}$ is downwards closed \label{itm:dcC}
\item For all $e\in F$, there exists a transition $X\xrightarrow{\underline{e}}$ \label{itm:Ftrans}
\item For $X_1,X_2,X_3\in \textsf{C}$: 
\begin{enumerate}
\item if $X_1\subseteq X_2\subseteq X_3$, $X_1\xrightarrow{A\cup\underline{B}}$, and $X_3\xrightarrow{A\cup \underline{B}}$, then $X_2\xrightarrow{A\cup \underline{B}}$ \label{itm:trans1}
\item if ${({(X_1\cup X_2)}\setminus {B})\cup A}\in {\textsf{C}}$, $X_1 \xrightarrow{A\cup\underline{B}}$, and $X_2 \xrightarrow{A\cup \underline{B}}$, then ${X_1\cup X_2} \xrightarrow{A\cup \underline{B}}$ and ${X_1\cap X_2}\xrightarrow{A\cup\underline{B}}$ \label{itm:trans2}
\item if $X_0,X_1,X_2,X_3\in \textsf{C}$, $A_0,A_1,B_0,B_1\subseteq E$ and there exist transitions $X_0\CStrans{A_0}{B_0}X_1$, $X_0\CStrans{A_1}{B_1}X_2$, $X_1\CStrans{A_1}{B_1}X_3$, and $X_2\CStrans{A_0}{B_0}X_3$, then $X_0\CStrans{A_0\cup A_1}{B_0\cup B_1} X_3$ \label{itm:trans3}
\end{enumerate}
\end{enumerate} 
\end{definition}

Figure \ref{fig:CS} shows a stable CS. One way to make it not stable would be to remove the transition from $\emptyset$ to $\{a,b\}$, since that would violate Item \ref{itm:trans3}.

As \cite{journals/jlp/PhillipsU15} did for RPESs and RAESs, we define a subcategory of \emph{cause-respecting} RESs in Definition~\ref{def:ccSRES}. This is based on the idea that if $e'$ enables $e$, then $e'$ cannot be reversed from a configuration which does not have another possible enabling set for $e$. Unlike causal reversibility \cite{danos2004reversible} however, a configuration fulfilling these conditions does not guarantee that reversing is possible.

\begin{definition}[Minimal enabling configurations for RES $m_{RES}(e)$]
Given an RES $\M{E}=(E,\textsf{Con},\vdash)$ the set of minimal enabling configurations of an event $e\in E$ is defined as:

$m_{RES}(e)=\{X\in \textsf{Con} \mid \exists Y. X\ob Y\vdash e\text{ and } \forall X',Y'. X'\ob Y'\vdash e\Rightarrow X'\not\subset X\}$
\end{definition}

\begin{definition}[CRES]\label{def:ccSRES}
A CRES $\M{E}=(E,\textsf{Con},\vdash)$ is an RES such that for all $e,e'\in E$, $e'\in X\in m_{RES}(e)$ iff whenever $X'\ob Y'\vdash \underline{e'}$, we have $e\in Y'$ or there exists an $X''\subseteq X'\setminus\{e'\}$ such that $X''\in m_{RES}(e)$.
\end{definition}

Moreover we define a \emph{cause-respecting} CS in much that same way as a CRES (Definition \ref{def:ccSCS}). This has the property that if $\M{E}$ is a CRES then $C_r(\M{E})$ is a CCS, and if $\M{C}$ is a finitely enabled CCS then $R(\M{C})$ is a CRES. In addition, the functors $C_r$ and $R$ are inverses of each other (Theorem \ref{the:CSRESinv}).

We can then prove Theorem~\ref{the:fwdreach}, which is analogous to a property of cause-respecting RPESs and RAESs proved in \cite{journals/jlp/PhillipsU15}. The CS in Figure \ref{fig:CS} is cause-respecting, but removing  the transition from $\emptyset$ to $\{a\}$ would change that.

\begin{definition}[Minimal enabling configurations for CS $m_{CS}(e)$]
Given a CS $\M{C}=(E,F,\textsf{C},\rightarrow)$ the set of minimal enabling configurations of an event $e\in E$ is defined as \[m_{CS}(e)=\{X\in \textsf{C}\mid X\xrightarrow{\{e\}} \text{ and } \forall X'.X'\xrightarrow{\{e\}}\Rightarrow X'\not\subset X\}\]
\end{definition}
\begin{definition}[CCS]\label{def:ccSCS}
A cause-respecting CS $\M{C}=(E,F,\textsf{C},\rightarrow)$ is a CS such that if $e'\in X\in m_{CS}(e)$, then whenever $X'\xrightarrow{\{\underline{e'}\}} Y'$ and $e\in X'$, there exists an $X''\subseteq Y'$ such that $X''\in m_{CS}(e)$.
\end{definition}


\begin{proposition}
If $\M{E}$ is a CSRES then $C_r(\M{E})$ is a CSCS, and if $\M{C}$ is a CSCS then  $R(\M{C})$ is a CSRES.
\end{proposition}

\begin{theorem}\label{the:CSRESinv} Given a SCS $\M{C}=\CS{}$, $C_r(R(\M{C}))=\M{C}$ if $\textsf{C}$ is downwards closed, and for all $e\in F$ there exists a transition $X\xrightarrow{\underline{e}}$.
If $\M{E}=\RES{}$ is a SRES with no ``unnecessary'' enablings $X\ob Y'\vdash e^*$ such that $X\ob Y\vdash e^*$ for $Y\subset Y'$ then $R(C_r(\M{E}))=\M{E}$.
\end{theorem}

\begin{theorem}\label{the:fwdreach}
If $\M{C}=(E,F,\textsf{C},\rightarrow)$ is a CSCS then every reachable configuration is forwards reachable.
\end{theorem}

\section{Conclusion}\label{sec:con}
We have defined categories for configuration systems (CS), reversible prime event structures (RPES), reversible asymmetric event structures (RAES), and reversible general event structures (RES), and functors between them, showing all the event structures can be modelled as CSs and conversely finitely enabled CSs can be modelled as RESs in a way that preserves morphisms, with the two directions being inverses in the stable setting (Theorem~\ref{the:CSRESinv}). We also defined coproducts for each of these categories, though products only for RESs and CSs. 

With a view to the semantics of causal reversible process calculi, we have also defined stable and cause-respecting subcategories of RESs, in which every reachable configuration is forwards reachable (Theorem~\ref{the:fwdreach}).

\textbf{Future Work:}
Defining a product of RPESs and RAESs will likely be trickier than for RESs, since definitions of products of prime event structures are far more complex than those of general event structures \cite{vaandrager1989simple}, and we note that the product of asymmetric event structures is as yet undefined. 
We plan to formulate a notion of `causal' RES which strengthens
the `cause-respecting' safety condition with a liveness condition.

\textbf{Acknowledgements:} We thank the referees for their helpful comments. This work was partially supported by EPSRC DTP award; EPSRC projects EP/K034413/1, EP/K011715/1, EP/L00058X/1, EP/N027833/1 and EP/N028201/1; EU FP7 612985 (UPSCALE); and EU COST Action IC1405.

\bibliography{bib}
\bibliographystyle{eptcs}

\end{document}